\documentclass[preprint]{emulateapj}
\newcommand{\icarus}{Icarus}

\usepackage{natbib}
\begin{document}

\title{The 3-4 $\mu$m spectra of Jupiter Trojan asteroids}

\author{M.E. Brown}
\affil{Division of Geological and Planetary Sciences, California Institute
of Technology, Pasadena, CA 91125}
\email{mbrown@caltech.edu}
\begin{abstract}
To date, reflectance spectra of Jupiter Trojan asteroids have revealed no
distinctive absorption features. For this reason, the surface composition
of these objects remains a subject of speculation. Spectra have revealed,
however, that the Jupiter Trojan asteroids consist of two distinct 
sub-populations which differ in the optical to near-infrared colors. 
The origins and compositional differences between the two sub-populations
remain unclear. Here we report the results from a 2.2-3.8 $\mu$m 
spectral survey of a collection of 16 Jupiter Trojan asteroids, divided
equally between the two sub-populations. We find clear spectral 
absorption features centered around 3.1 $\mu$m
in the less red population. Additional absorption consistent
with expected from organic materials might also be present. No such
features are see in the red population. A strong
correlation exists between the strength of the 3.1 $\mu$m absorption
feature and the optical to near-infrared color of the objects. 
While 
traditionally absorptions such as these in dark asteroids are modeled
as being due to fine-grain water frost, we find it 
physically implausible that the special circumstances required to create 
such fine-grained frost would exist on a substantial fraction of the Jupiter
Trojan asteroids. We suggest, instead,
that the 3.1 $\mu$m absorption on Trojans and other dark asteroids could be 
due to N-H stretch features. 
Additionally,
we point out that reflectivities derived from WISE observations show
a strong absorption beyond 4$\mu$m for both populations. 
The continuum of 3.1 $\mu$m features and the common absorption beyond 4 $\mu$m
might suggest that both sub-populations of Jupiter Trojan asteroids formed
in the same general region of the early solar system.
\end{abstract}
\section{Introduction}
The Jupiter Trojan asteroids remain mysterious in both their origins
and in their compositions. As small bodies
 orbiting along the inner boundary of the
realm of the giant planets, their histories contains important 
clues to the origins and evolution of the entire planetary system.

Recent proposals have suggested that the Jupiter Trojans were emplaced from
the outer solar system during an early period of giant planet dynamical
instability \citep{2005Natur.435..462M, 2013ApJ...768...45N}. 
In this hypothesis, the objects that 
are now Jupiter Trojans and the objects that are now in the Kuiper belt 
are derived from the same source populations but underwent different stochastic
dynamical histories to lead to their very separate locations today. 
Jupiter Trojans are, in this view, a much more easily accessed 
population of the types of
icy bodies inhabiting the outer solar system. 

While the dynamical connection between the Jupiter Trojan and Kuiper belt
source region seems promising, little progress has been made in 
connecting the compositions of these two sets of objects. Ices
such as water and methanol have been seen on small objects in the
Kuiper belt \citep{2011Icar..214..297B,2012AJ....143..146B}, but
reflectance spectroscopy of
Jupiter Trojans has revealed nothing but featureless red spectra.
These red spectra have typically been interpreted as being due to
the presence of organic materials, but no such spectral signatures
have ever been seen. Indeed, the only positively identified spectral
features on Jupiter Trojans comes from thermal emission spectra, which
show the signatures of fine-grained silicates \citep{2006Icar..182..496E}. 
Spectral
models including reddened space-weathered silicates appear to
explain the visible-to-near-infrared spectra as effectively as any organic
materials \citep{2011AJ....141...25E}.

An important clue into the compositions of the Jupiter Trojans and 
perhaps into their connection with objects in the Kuiper belt is the
insight that the Jupiter Trojans are a mixture of two 
populations with distinct (though still featureless) spectra \citep{2011AJ....141...25E}. 
The ''red'' and ''less red'' populations differ in their optical and
near-infrared
colors, their infrared reflectivities measured from the WISE spacecraft,
and in their size distributions \citep{2011AJ....141...25E, 2012ApJ...759...49G,2014AJ....148..112W,2015AJ....150..174W}, yet seem
completely mixed dynamically. It is
not known whether these two color populations represent
distinct source populations, distinct dynamical pathways,
distinct collision histories, or some other set of mechanisms.
To date, the only hypothesis for the two populations is that 
they are both from the Kuiper belt source region, but the less-red
objects formed in the inner part of the source region where H$_2$S 
evaporated from the surface, while the red objects retained their surface
H$_2$S \citep{ian}. In this hypothesis, subsequent irradiation then caused
 significant chemical
differences in the sulfur-free and sulfur-containing involatile
mantles \citep{2016ApJ...820..141M}, leading to the bifurcation of not just
Jupiter Trojan colors, 
but also to those of the Centaurs \citep{2008ssbn.book..105T} and the small Kuiper belt objects
\citep{2012ApJ...749...33F,2015DPS....4720303W}

To further explore the compositions of the Jupiter Trojans,
to examine connections with the Kuiper belt population, and
to determine differences between the two populations, we have
obtained 2.2-3.8 $\mu$m spectra of a sample of 8 of the less-red and 8 of the red
Jupiter Trojans. This spectral region is potentially fruitful as ices which
are seen in Kuiper belt objects have some of
their strongest fundamental absorptions at these wavelengths and
because dark outer main belt asteroids have been found to show some
of their most distinctive absorption features in these regions
\citep{2012Icar..219..641T}. 

\section{Observations and data reduction}
We observed 16 Jupiter Trojan asteroids over a total of 8 nights
from 2013 to 2015 using the facility NIRSPEC infrared medium resolution
spectrograph at the Keck Observatory (Table 1). 
The observing procedure and data reduction
were performed identically to the observations described in \citet{2014ApJ...793L..44B}.
In short, the target was identified by its position in the sky and by its 
motion with respect to background stars in short imaging exposures. 
The target was then placed on a 0.57 arc second wide 
spectral slit for a long series of 
nodded exposures. During this series of exposures, guiding on the target
was performed using the infrared slit-viewing camera. We required two
separate instrument settings to cover the range from 2.2 to 3.8 $\mu$m.
At the shorter wavelength setting, which covered from approximately
2.2 to 3.1 $\mu$m at a median
(2 pixel) spectral resolution, $\lambda / \Delta \lambda$, of 1600,
we obtained between 18 and 40 nodded exposures,
typically with 10 second integrations and 15 second coadds. For
the longer wavelength setting, which covered from approximately 
3.0 to 3.8~$\mu$ at a
median spectral resolution of 2000, we
obtained between 16 and 46 nodded exposures, typically with 1 second
integrations and 150 second coadds. Beyond 3.3 $\mu$m, thermal drifts
occasionally prevented high quality background subtraction. All spectra
were checked by eye, and those with poor background subtraction were
discarded. Correction for telluric absorption features was performed by
dividing the extracted spectrum by the spectrum of a nearby solar-type
star. For the two faintest Jupiter Trojans in the survey (objects
659 and 4060), the longer
wavelength setting yielded no usable data, so only the short wavelength
setting is used.
\section{Results}
We first examine the brightest red and brightest less-red objects in our sample.
Hektor, the largest Jupiter Trojan, is also the reddest object in our sample.
Figure 1 shows the 2.2 - 3.8 $\mu$m spectrum of (624) 
Hektor along with a continuum
extrapolation that was estimated by taking the $H-K$ synthetic photometry
from \citet{2011AJ....141...25E} and extending it linearly out to 4 $\mu$m. Hektor appears
featureless to the level of the signal-to-noise, and the extrapolation from
shorter wavelengths provides an excellent prediction of the continuum
level. Patroclus, the brightest less-red object in our sample (Figure 2), 
in contrast, does not follow this trend. At
3 $\mu$m the flux is $\sim$10\% below the extrapolated continuum, and it
rises well above the continuum by 3.8 $\mu$m. Smaller scale features are
potentially visible, but we first chose to examine the entire data set before
focusing on these features. 
\begin{figure}
\plotone{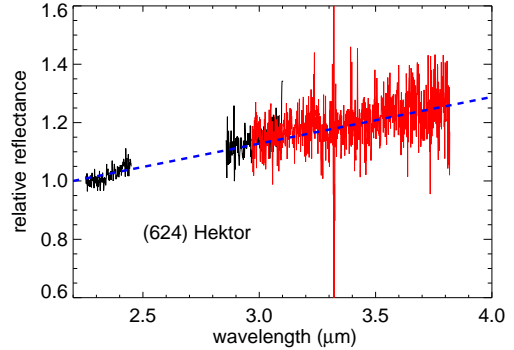}
\caption{The largest Jupiter Trojan, (624) Hektor, is featureless in
this wavelength range. The black
and red lines show the two separate spectral settings used to cover this
wavelength range, with an overlap between 2.85 and 3.1 $\mu$m.
The dashed blue line is an extrapolation from the K-band spectrum of
\citet{2011AJ....141...25E}.}
\end{figure}
\begin{figure}
\plotone{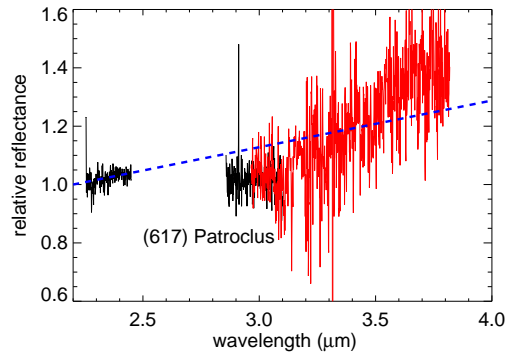}
\caption{The largest less red Jupiter Trojan, (617) Patroclus, has
a distinct absorption beyond 3 $\mu$m.  The black
and red lines show the two separate spectral settings used to cover this
wavelength range, with an overlap between 2.85 and 3.1 $\mu$m.
The dashed blue line is an extrapolation from the K-band spectrum of
\citet{2011AJ....141...25E}.}
\end{figure}

To highlight the broadest features of the spectra, we degrade the spectral
resolution
by a factor of 8 for all of the spectra by convolving each spectrum with
a gaussian function with a full-width-half-maximum of 16 pixels and sampling
every 8 pixels, simulating Nyquist sampled spectra with resolutions of
200 and 250 in the short and long wavelength settings, respectively. 
Figure 3 shows these spectra for all objects, along with the 
extrapolation from their $H-K$ photometry. The data show that Hektor and
Patroclus appear to represent two different types of
spectra. The Hektor-like spectra all appear generally featureless and lie
approximately along the linear extrapolation from the H-K continuum, while
the Patroclus-like spectra come to a minimum at approximately 3.1 $\mu$m
and rise sharply out to 3.8 $\mu$m.
\begin{figure}
\plotone{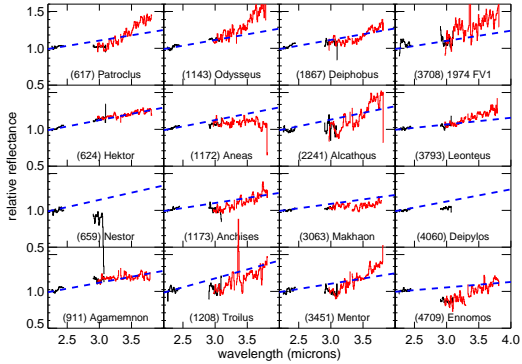}
\caption{All spectra, smoothed in resolution by a factor of 8 from the 
original data. Some spectra appear featureless like Hektor,
while others show spectral features similar to that seen on
Patroclus. }
\end{figure}

To examine the relationship between the 3 $\mu$m absorption seen on 
some Jupiter Trojans and the two Jupiter Trojan spectral types, we
construct a simple measure of the depth of the 3.1 $\mu$m feature
by linearly interpolating a predicted 3.1 $\mu$m flux between the median
of the data from 2.25 to 2.45 $\mu$m
 and the median of the data beyond 3.5 $\mu$m and then
subtracting this interpolation from the measured median of the data between
3.05 and 3.15 $\mu$m. 
In Figure 4, we compare this 3.1 $\mu$m absorption depth
to the [.85-J] color from \citet{2011AJ....141...25E}, which they show is 
an excellent discriminator of the two spectral classes. For two
objects, no spectral data are available beyond 3.1 $\mu$m. We find,
however, that the 3.1 $\mu$m absorption depth is strongly correlated
with a simple subtraction of the median of the data from 2.25 to 2.45 $\mu$m from the median of the data from 2.86 to 3.1 $\mu$m. We estimate 
the uncertainties in this estimate by taking
the full range of 3.1 $\mu$m absorption depths measured from
objects with the same 2.3 to 3.0 $\mu$m difference.
\begin{figure}
\plotone{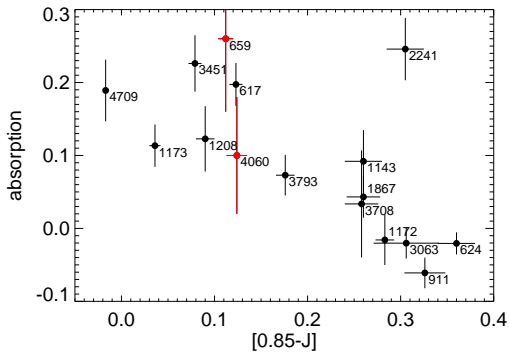}
\caption{The 3.1 $\mu$m absorption depth strongly correlates with the
[.85-J] color for Jupiter Trojans. The less red object (defined to be
those with [.85-J]$<0.2$ all show absorption, with some of the strongest
absorption on the least red objects. The red objects generally show
little to no absorption, with (2241) Alcathous as a significant outlier.
The objects show in red, (4060) and (659), have spectral data only out to
3.2 $\mu$m and their absorption depths have been estimated from the 2.4
to 3.1 $\mu$m ratio.}
\end{figure}

The correlation between the [0.85-J] color and 3.1 $\mu$m absorption depth
is striking. A Spearman rank correlation test shows that the probability
of such a high correlation due to chance is only 0.3\%.
Interestingly, the 3.1 $\mu$m absorption depth appears more correlated with
[0.85-J] color than simply bifurcated into a red and less-red group. 
Three objects of the red group -- (3707), Deiphobus, and Odysseus --
have features which appear similar to the long wavelength portion of
Patroclus, while one of the red objects -- Alcathous -- is a clear
outlier with one of the strongest 3.1 $\mu$m absorptions. The 
presence of a similar looking
3.1 $\mu$m absorber in both populations might be an
important clue into the relationship between the less red and red populations.

\section{Spectral modeling}
In an attempt to understand the compositional implications of 
the 3.1 $\mu$m absorption, we increase the signal-to-noise by taking an
average of the six spectra that show absorption depth greater than
10\%.
Figure 5 shows that the average of these spectra has a
clear absorption with a peak depth at
about 3.08 $\mu$m, with reflectivity rising at longer wavelengths.
A potential set of absorptions is also visible from 3.2 to 3.6 $\mu$m,
depending on where the continuum level is defined.
\begin{figure}
\plotone{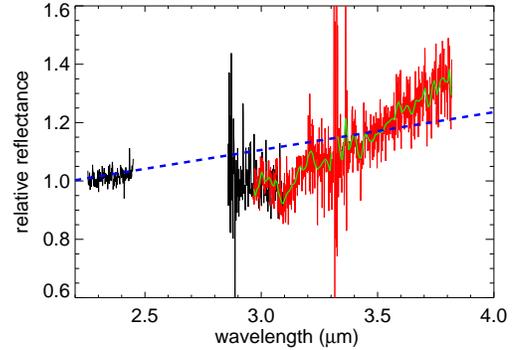}
\caption{The average spectrum of the six objects with strong 3.1 $\mu$m
absorption shows strong absorption from 2.9 to 3.2 $\mu$m and potential
evidence of weaker absorptions beyond. The dashed line shows the averaged
extrapolation of the six objects' K-band spectra from \citet{2011AJ....141...25E}.
The solid green line shows the average of the smoothed spectra from Figure 3.}
\end{figure}

In dark asteroids, 3 $\mu$m absorption bands are often attributed to 
OH bands of phyllosilicates or to water ice bands \citep{2010Natur.464.1322R, 
2010Natur.464.1320C,
2012Icar..219..641T}.
The decrease in reflectivity from 2.9 to 3.08 $\mu$m is the opposite of
the OH phyllosilicate bands which peak in absorption in the middle
of the unobserved atmospheric band and have steeply rising reflectivites from 2.9 $\mu$m onward (see the ``steep'' spectral type of Takir \& Emery).
Rather, the absorption feature appears similar to that seen in Themis
\citep{2010Natur.464.1322R, 2010Natur.464.1320C} and also in a small number of additional
dark outer main belt asteroids (the ``rounded'' spectral type of 
Takir \& Emery). On Themis, this feature has been attributed to fine 
grained frost covering dark silicate grains; path lengths through the
water grains must be quite short in order to have such a weak absorption
at 3.1 $\mu$m. Our spectra can be fit with a similar model; 
we use an identical
procedure to that used by \citet{2010Natur.464.1322R} to model Themis
and construct a Shkuratov model 
\citep{1999Icar..137..235S}
of dark grains covered
with fine grained ice.
The dark grains have optical constants chosen
to match the albedo level of the continuum
and the ice is a fine-grained frost
with a volume fraction of 1.1\% (chosen to match the depth of the absorption)
coating the dark grains. Water ice optical
constants are obtained from \citet{2009ApJ...701.1347M}.
The spectral fit to the
3.1 $\mu$m band is quite good (Figure 6). The continuum continues to rise at
longer wavelengths, however, and broad absorption from 3.2 to 3.6 $\mu$m 
may also be present. 
\begin{figure}
\plotone{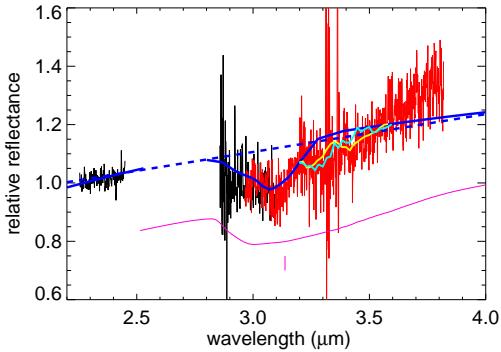}
\caption{The 3.1 $\mu$m region of the spectrum of the Jupiter Trojans 
with features in this wavelength range can be fit by a model of
dark grains covered by fine-grained water ice frost (dark blue),
though we do not find such a composition to by physically plausible.
The magenta line offset downward shows a laboratory spectrum of
the involatile residue after the irradiation of N$_2$+CO+CH$_4$ ices. 
While much of the spectrum is dominated by water contamination, 
an absorption peak due to an N-H stretch, 
marked, appears at 3.13 $\mu$m in a location
similar to the peak of the Trojan absorption. Regardless of the interpretation
of the 3.1 $\mu$m region,
additional absorption appears beyond 3.2 $\mu$m. These absorption
appears similar to those attributed to aromatic and aliphatic hydrocarbons
on Saturnian irregular satellites. Phoebe (light blue) shows primarily
the 3.3 $\mu$m aromatic feature, while Hyperion (yellow) shows a
stronger 3.4 $\mu$m aliphatic feature.}
\end{figure}

Though the fit to fine-grained water ice frost is good, we are skeptical
of this interpretation for Trojan asteroids. Extremely fine grained frost
is required in order to show a modest 3$\mu$m absorption while showing no
hints of the shorter wavelength water ice features. 
While the idea that Themis,
a known active asteroid, might have water vapor continuously outgassing
from the interior and temporarily creating a fine grained frost
seems plausible, such an interpretation for
Jupiter Trojans asteroids
seems physically unlikely. 
\citet{2014Icar..231..232G} shows that water ice should be
depleted to depths of $\sim$10m in typical Jupiter Trojan asteroids,
so unusual circumstances like recent collisions would be required to
have substantial active sublimation. Having such unusual circumstances
in a large fraction of the Jupiter Trojan asteroids seems unlikely.
We instead seek a simpler explanation for the observed absorption
that does not require special circumstances.
We suggest that the 3.1 $\mu$m absorption feature could
be due to an N-H, rather than O-H stretch. Laboratory experiments of
outer solar system ices which include nitrogen show the creation of
residues with spectra similar to poly-HCN and absorptions at
about 3.1 $\mu$m \citep{2014ApJ...788..111M, 2015ApJ...812..150M}.
Figure 
6 shows a spectrum of electron irradiated N$_2$+CH$_4$+CO 
from \citet{2015ApJ...812..150M}, and while
much of the 3$\mu$m region is dominated by water absorption (a contaminant
in the laboratory measurement), an absorption peak at 3.13$\mu$m due to
N-H appears at nearly the same location as the Trojan absorption. The
laboratory experiments were not designed to specifically investigate
Jupiter Trojan asteroid physical conditions and no optical constants are available to
facilitate more quantitative modeling, so the lack of a perfect match is
unconcerning.
In the context of dynamical instability models,
nitrogen should indeed be present on the surface of small
bodies beyond the giant planets in the form of NH$_3$ so it should be 
expected in the involatile surface mantles upon
irradiation \citep{ian}. If the H$_2$S evaporation
hypothesis of \citep{ian} is correct, the lack of 3.1 $\mu$m absorption
in most red Jupiter Trojan asteroids would mean that the
presence of sulfur must disrupt much of the N-H
chemistry. Such a possibility is amendable to laboratory exploration
using experiments such as those performed in \citep{2016ApJ...820..141M}.

Regardless of the spectral assignment,
no known asteroid has this particular set of spectral features, though the
asteroid Themis has both a $\sim$3.1$ \mu$m absorption
 and a set of longer
wavelength absorptions that have been attributed to organic materials \citep{2010Natur.464.1322R, 2010Natur.464.1320C}. 
To search
for possibly better analogs to Jupiter Trojans, we examine
instead bodies from the outer solar system. The Jovian irregular 
satellite Himalia has a visible-to-4 $\mu$m spectrum quite unlike any
of the Jupiter Trojans, including a broad 1-2 $\mu$m 
absorption and a (52) Europa-like
3 $\mu$m absorption of unknown origin \citep{2014ApJ...793L..44B}.  
The dissimilarity between the Jupiter Trojans and Himalia is
surprising, at least in the context of instability models which predict
they come from the same outer solar system source population. The
Jovian irregular satellites are in a significantly more intense collisional
environment, however, so perhaps differences can be attributed to
the effects of impacts. If so, this interpretation may have interesting
implications for the other objects with (52) Europa-like spectra that
reside in the asteroid belt.

The irregular satellites of the Saturnian system appear more promising.
Phoebe has absorption features centered near
3.3 and 3.4 $\mu$m which have been attributed to aromatic and aliphatic
hydrocarbons, respectively \citep{2014Icar..233..306C}. 
These features appear similar to
those on these Jupiter Trojans asteroids. In Figure 6 we add to our
spectral model a continuum-subtracted spectrum of Phoebe from 3.2 to
3.6 $\mu$m. The match to the 3.3 $\mu$m absorption is good,
while Phoebe's 3.4 $\mu$m absorption is not as deep as the potential
feature on the Jupiter Trojans. Iapetus and Hyperion also show
these spectra features, and on these the 3.4 $\mu$m feature is a closer
match to that on the Jupiter Trojans. While the signal-to-noise is
insufficient to definitively identify these features as 
aromatic and aliphatic hydrocarbon absorption features, the match with
the features on the Saturnian satellites is notable.

An unusual feature of these spectra, and one that can be clearly seen 
even in the individual spectra of Figure 3, is that the extrapolated
continuum under-predicts the reflectance beyond 3.6 $\mu$m. Such behavior
is not seen in the spectra of any of the dark asteroids nor in any of
the small bodies in the outer solar system for which spectra are available.
While this rise in flux at longer wavelengths appears similar to that 
expected from thermal emission, at the low temperatures of the Jupiter
Trojans the thermal flux is still a small fraction of the emission at
these wavelengths, and, as seen below, the flux drops at longer wavelengths. 

The same reflectivity rise beyond 3.6 $\mu$m is seen in the red objects
which are categorized as having a 0-10\% 3.1 $\mu$m absorption in Figure 3. 
These objects do not show a distinct shorter wavelength
 band, but appear otherwise
similar to the less red objects beyond 3.6 $\mu$m. These objects also
contain a few tantalizing features in the 2.9 to 3.1 $\mu$m range, 
but we deem none of them sufficiently reliable to seriously examine.
The most red of the 
red objects, however, show no 3.1 $\mu$m
absorption and no rise beyond 3.6 $\mu$m.

To further explore the complete spectrum of the Jupiter Trojans, we
construct mean absolute reflectances
 from 0.4 to 5 $\mu$m for the two spectral types (Figure 7).
For the spectrum from 0.4 to 2.5 $\mu$m we use the mean red and less
red spectra of
\citet{2011AJ....141...25E}
scaled to the mean measured visible albedos from
\citet{2012ApJ...759...49G}. From
2.2 to 3.8 $\mu$m we use our average spectra for the red and less-red
spectral types, scaled to match the small 2.2 to 2.5 $\mu$m overlap region.
For longer wavelengths, we use average WISE reflectivities from \citet{2012ApJ...759...49G}. 
The WISE observations bifurcate into the same red and less-red spectral
types \citep{2014AJ....148..112W}, and we use all observed
objects with WISE calculated diameters greater than 80 km (at smaller
diameters the uncertainties begin to dominate the measurements).
We apply
no scaling to the WISE reflectivities, 
but simply take the average absolute reflectivities
measured in the 2.9 to 3.7$\mu$m W1 band and in the 4.0 to 5.1 $\mu$ W2 band.
The precise match between our extrapolated spectral reflectivities and 
the measured WISE reflectivities in the W1 band region is encouraging.
\begin{figure}
\plotone{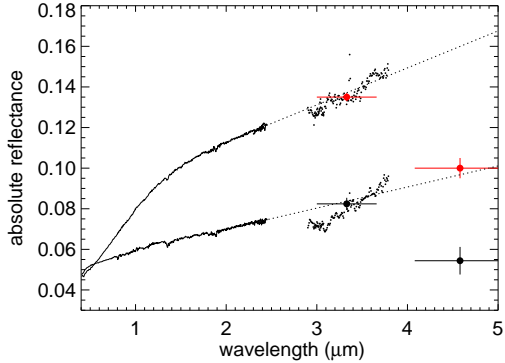}
\caption{The average of the 0.4 to 5.0 $\mu$m spectral of the two
spectral classes shows similarities and differences throughout
the spectral range. The 0.4 to 2.4 $\mu$m data come from \citet{2011AJ....141...25E},
the 2.2 to 3.8 $\mu$m data are from this paper, and the two 
broad spectrophotometric points are from WISE photometry \citep{2012ApJ...759...49G}.
The two red points show WISE photometry of red objects, while the two
black points show WISE photometry of less red objects. No scaling is 
performed between the WISE data and our data; the match in the 3.5 $\mu$m
band is excellent. The sharp drop in reflectivity for both the red
and the less red types is the most significant feature seen in this
wavelength range. Much of the structure in the 3 - 4 $\mu$m region
of the red spectral type comes from adding in the single outlier (2241)
Alcathous.}
\end{figure}

The sharp drop in reflectivity between the WISE W1 and W2 bands 
is perhaps the most surprising feature in this spectral range. 
The mean W2 to W1 reflectance ratio of all Jupiter Trojans larger than 80 km
is 0.72$\pm$0.03. For both
classes of objects, the drop in reflectivity compared to the extrapolated
continuum is at least a factor of two over this $\sim$1 $\mu$m range,
suggesting a broad and deep absorption. Such a broad and deep absorption in 
4 to 5 $\mu$m range with no corresponding feature at shorter wavelengths
has not yet been seen on any observed body in the solar system. 
Phoebe, for example,
has a broad absorption at these wavelengths \citep{2008Icar..193..309B}, but it
is due to water ice which has a very strong band at 3 $\mu$m. On these
Jupiter Trojans, the at most very small water band at 3 $\mu$m predicts
that no band will be detectable at 4.5 $\mu$m. The object (32532) Thereus
is the only Centaur detected in both the W1 and W2 bands by WISE. 
The reflectivity in the W2 band is a factor of 2 {\it higher} than that
in the W1 for this object \citep{2013ApJ...773...22B}. The 3 Jovian
irregular satellites that were observed in the W1 and W2 bands have a mean
W2 to W1 reflectivity ratio of 1.2$\pm$0.4, which has too large of 
an uncertainty to allow a useful conclusion to be drawn \citep{2012ApJ...759...49G}.

One possibility for the low reflectivity from 4 to 5 $\mu$m is that it
is caused by silicates. \citet{1995A&A...300..503D} find that Mg rich laboratory 
manufactured pyroxenes glasses have spectra that 
can drop sharply in reflectivity beyond 4 $\mu$m. The laboratory glasses
were constructed to be analogues of interstellar grains and are
also representative of dust grains expected in cometary comae. Such silicate
grains could also be the species responsible for the 10 $\mu$m emission
detected by \citet{2006Icar..182..496E} from Spitzer Space Telescope spectroscopy.
Spectral modeling by \citet{2011AJ....141...25E} shows that such Mg rich pyroxenes can
also explain much of the red coloration of these objects. 

\section{Discussion}
Jupiter Trojan asteroids exhibit a range of 2.2 to 3.8 $\mu$m spectra
strongly correlated with their optical colors. 
The least-red objects (and one red outlier) have distinct 3.1 $\mu$m
absorptions that can be modeled as a small covering a fine-grained water
ice but is also consistent with N-H stretch features seen in irradiated
outer solar system ice experiments.
Additional features similar to those seen and attributed to organic
materials on Saturnian irregular satellites are also present. 
The full visible to 3.8 $\mu$m
spectra most closely resemble the ``rounded'' 3 $\mu$m spectral type
of \citet{2012Icar..219..641T}. In particular, 
asteroids such as (153) Hilda, the largest
P-type asteroid in the Hilda group, have spectra which are indistinguishable
from these less-red Jupiter Trojans. The spectrum of Hilda does not
have sufficient signal-to-noise beyond 3.3 $\mu$m to confirm the presence
of the potential organic absorptions, but their presence cannot be ruled out.

The red Jupiter Trojans (with the exception of the single outlier)
contain no detectable 3.1 $\mu$m absorption feature and no
detectable organic features, but instead simply a red optical spectrum
that flattens to a featureless infrared spectrum. No such spectra have
been seen in the asteroids nor in the irregular satellites.
The colors of the red Jupiter Trojans are indistinguishable from
the colors of the red Kuiper belt objects \citep{2008ssbn.book..105T}, yet these Kuiper belt
objects contain sufficient amounts of water ice on their surfaces
\citep{2011Icar..214..297B,2012AJ....143..146B}
that they would be easily detectable as strong 3 $\mu$m absorptions
on Jupiter Trojans. 

The interpretation of the 3.1$\mu$m absorption seen on the less-red
Jupiter Trojans as being due to an N-H band is broadly consistent
with the idea that these objects formed beyond the giant planet region,
where NH$_3$ would have been stable, and were irradiated to form involatile
residues before being emplaced at 5 AU. \citet{ian} propose that the
red Jupiter Trojans formed in the outer part of the original outer disk
and would have had H$_2$S stable on the surface during irradiation, leading
to much redder colors. In this hypothesis sulfur must suppress some
of the N-H chemistry, a suggestion that can be experimentally tested 
and, if verified, would provide strong evidence that all Jupiter Trojan
asteroids formed beyond the giant planet region and that the differences
in the two color populations are only due to surficial differences 
in ice irradiation.

An additional suggestion that the two spectral types are compositionally
related comes from the observation that
they both have reflectivities that drop sharply beyond 4 $\mu$m.
Few species are capable of providing such a strong absorption 
exclusively beyond these wavelengths, so it seems highly likely
that the material causing this reflectivity drop is the same in 
both the red and less-red Jupiter Trojans.

Two key questions about Jupiter Trojans asteroids remain: (1) Did the
two spectral classes originate from the same region? (2) Did they originate
from beyond, within, or inside of the giant planet region? 
These data suggest the possibility that the answer to the first question
is yes. Though the two populations of Jupiter Trojans asteroids are
bifurcated in color, their 3-4 $\mu$m spectra form more of a continuum.
Moreover, both populations contain an unusual deep spectral absorption
between 4 and 5 $\mu$m. The second question remains unanswered, but 
confirmation of N-H absorption in the less-red objects along with verification that the presence of sulfur can mute the strength of the N-H absorption
would provide powerful evidence that these objects formed in the outer solar
system straddling the H$_2$S evaporation line, as suggested 
by the hypothesis of \citet{ian}.

\acknowledgements
We thank Andy Rivkin and Josh Emery for illuminating conversations about the
nature of 3 $\mu$m absorptions in dark asteroids. Discussions with the 
Keck Institute for Space Studies ``In Situ Science and Instrumentation
for Primitive Bodies'' study group, including Jordana Blacksberg, Bethany Ehlmann, John Eiler, Robert Hodyss, Ahmed Mahjoub, Michael Poston and Ian Wong
have been extremely valuable to the interpretation of these
data.  This research was supported by Grant NNX09AB49G from
the NASA Planetary Astronomy Program.
The data presented herein were obtained at the W.M. Keck Observatory, 
which is operated as a scientific partnership among the California Institute 
of Technology, the University of California and the National Aeronautics and 
Space Administration. The Observatory was made possible by the generous 
financial support of the W.M. Keck Foundation. The author wishes to 
recognize and acknowledge the very significant cultural role and reverence 
that the summit of Mauna Kea has always had within the indigenous 
Hawaiian community.  We are most fortunate to have the opportunity to 
conduct observations from this mountain.

\begin{deluxetable}{lccccc}
\tablecaption{Log of observations}
\tablehead{\colhead{object} & \colhead{date} & \colhead{spectral range} & \colhead{exp. time} & \colhead{airmass} & \colhead{calibrator} \\ 
\colhead{} & \colhead{(UT)} & \colhead{$\mu$m} & \colhead{(sec)} & \colhead{} & \colhead{}\\ } 
\startdata
(617) Patroclus & 23 Nov 2013 & 2.2-3.1 & 4500 & 1.00-1.02 & HD12846 \\
                &             & 3.0-3.8 & 4000 & 1.00-1.02 & HD12846 \\
(624) Hektor    & 16 Mar 2014 & 2.2-3.1 & 2700 & 1.08-1.18 & LHS2524 \\
                &             & 3.0-3.8 & 4200 & 1.20-1.52 & LHS2524 \\
(659) Nestor	& 04 May 2015 & 2.2-3.1 & 5500 & 1.09-1.24 & HD106116 \\
(911) Agamemnon & 02 May 2015 & 2.2-3.1 & 3600 & 1.56-1.63 & HD111564\\
		& 01 May 2015 & 3.0-3.8 & 6900 & 1.56-1.70 & HD111564\\ 
(1143) Odysseus & 01 May 2015 & 2.2-3.1 & 3600 & 1.24-1.52 & HD111031 \\
		& 	      & 3.0-3.8 & 4050 & 1.85-1.39 & HD111031 \\ 
(1172) Aneas    & 23 Nov 2013 & 2.2-3.1 & 1800 & 1.21-1.37 & LTT10989 \\
		&             & 3.0-3.8 & 2400 & 1.00-1.04 & LTT10989 \\
(1173) Anchises & 23 Nov 2013 & 2.2-3.1 & 2400 & 1.01-1.02 & HD9224 \\
		&             & 3.0-3.8 & 3750 & 1.00-1.12 & HD9224 \\
(1208) Troilus  & 26 Nov 2013 & 2.2-3.1 & 2700 & 1.00-1.03 & LTT10989 \\
		&             & 3.0-3.8 & 3150 & 1.00-1.08 & LTT10989 \\
(1867) Deiphobus& 25 Nov 2013 & 2.2-3.1 & 2700 & 1.17-1.22 & HD664 \\
		&             & 3.0-3.8 & 6450 & 1.00-1.11 & HD664 \\
(2241) Alcathous& 23 Nov 2013 & 2.2-3.1 & 1000 & 1.00-1.01 & LHS1013\\
		& 25 Nov 2013 & 2.2-3.1 & 2700 & 1.06-1.14 & LHS1013 \\
		& 23 Nov 2013 & 3.0-3.8 & 2550 & 1.02-1.13 & LHS1013 \\
		& 24 Nov 2013 & 3.0-3.8 & 1350 & 1.04-1.16 & LHS1013 \\
(3063) Makhaon  & 02 May 2015 & 2.2-3.1 & 5400 & 1.50-1.54 & HD132173 \\
		& 03 May 2015 & 3.0-3.8 & 4200 & 1.49-1.62 & HD132173\\
(3451) Mentor 	& 24 Nov 2013 & 2.2-3.1 & 2700 & 1.20-1.30 & HD27466 \\
		&	      & 3.0-3.8 & 2100 & 1.11-1.13 & HD27466\\
		& 26 Nov 2013 & 3.0-3.8 & 2850 & 1.15-1.37 & HD27466 \\
(3708) 1974 FV1 & 26 Nov 2013 & 2.2-3.1 & 3600 & 1.21-1.51 & LHS1428\\
		&	      & 3.0-3.8 & 2700 & 1.03-1.16 & LHS1428 \\
(3793) Leonteus & 01 May 2015 & 2.2-3.1 & 4200 & 1.22-1.61 & HD133352 \\
		& 02 May 2015 & 3.0-3.8 & 6000 & 1.14-1.54 & HD133322 \\
(4060) Deipylos & 02 May 2015 & 2.2-3.1 & 5400 & 1.50-1.54 & HD132173 \\
(4709) Ennomos  & 25 Nov 2013 & 2.2-3.1 & 5400 & 1.10-1.36 & LTT10989 \\
		&             & 3.0-3.8 & 4050 & 1.01-1.07 & LTT10989 \\
 
\enddata
\end{deluxetable}

\end{document}